\documentclass[useAMS,usenatbib]{mn2e-mod}

\usepackage{epsfig}
\usepackage{graphicx}
\usepackage{url}

\title[Bug in Mercury]{Symplectic Integrator Mercury: Bug Report}
\author[K. de Souza Torres and D. R. Anderson]{K. de Souza Torres$^{1}$\thanks{E-mails:
karlchen79@gmail.com (KST); dra@astro.keele.ac.uk (DRA)} and D. R.
Anderson$^{2}$\footnotemark[1]\\
$^{1}$Instituto Nacional de Pesquisas Espaciais - INPE, S\~ao Jos\'e dos Campos, 12211090. Brazil\\
$^{2}$Astrophysics Group, Lennard-Jones Laboratories, Keele University, Staffordshire ST5 5BG, UK}
\begin{document}


\pagerange{\pageref{firstpage}--\pageref{lastpage}} \pubyear{2008}

\maketitle

\label{firstpage}

\begin{abstract}
We report on a problem found in {\sc Mercury}, a hybrid symplectic integrator used for dynamical problems in Astronomy. The variable that keeps track of bodies' statuses is uninitialised, which can result in bodies disappearing from simulations in a non-physical manner. Some {\sc fortran} compilers implicitly initialise variables, preventing simulations from having this problem. With others compilers, simulations with a suitably large maximum number of bodies parameter value are also unaffected. Otherwise, the problem manifests at the first event after the integrator is started, whether from scratch or continuing a previously stopped simulation. Although the problem does not manifest in some conditions, explicitly initialising the variable solves the problem in a permanent and unconditional manner.
\end{abstract}


\section{Introduction}
\label{intro}
{\sc Mercury} \citep{Chambers1999} is a general-purpose software package, 
written in {\sc fortran77}, for doing N-body integrations to investigate dynamical problems in Astronomy. In a set of simulations to study accretion dynamics using {\sc Mercury} \citep{Torres2008}, it was found that the results contained fewer planets than expected. Analysis of output files and creation of scripts and movies to carefully follow the processes showed discontinuities in the number of embryos during the integrations. No events for these bodies were 
being registered in the output files and their disappearance was non-physical. 

\section{The Problem}
\label{problem}
In {\sc mercury}, an integer variable array called {\sc stat} keeps track of the statuses of particles: whether they are alive, have been involved in a collision or have been ejected from the system. At the beginning of an old or new integration, all the bodies must have its {\sc stat} equals to zero (Table~\ref{stat}). When a live body is collided or ejected, its {\sc stat} value is changed accordingly and the body is removed by the subroutine called {\sc mxx\_elim}.

In the code, the {\sc stat} array is not initialised, meaning its elements may access random values. If these values are negative, not only bodies that have been involved in an event are going to be removed, but also those with invalid negative values in its respective {\sc stat} elements, causing non-physical disappearance of some bodies during the integrations. 

\begin{table}
\centering
\caption{Valid values for the variable STAT in Mercury integrator.}
\begin{tabular}{@{}cc@{}} 
\hline
{\sc stat} value & Body status\\
\hline
  0	&	Alive\\
 -2	&	Collided\\
 -3	&	Ejected\\
\hline
\label{stat}
\end{tabular}
\end{table}

\subsection{Characterising the problem}
\label{charac}
Not all simulations are affected by this problem. Some {\sc fortran} compilers (e.g. \textit{ifort}) implicitly initialise variables on their declaration statement \citep{Fortran}. Using one of them, an integer array would have all its elements set to zero in its declaration, even without an explicit initialisation in the source code. Others compilers (e.g. \textit{g77} and \textit{gfortran}) begin initialising the early elements of an integer array with zero when the array is beyond a certain size; this size will be compiler and machine environment dependent. With these compilers, the results will be affected by the non-initialisation of the {\sc stat} array if uninitialised elements are used; it is this we are about to investigate further.

The {\sc stat} array has its size initially defined by the parameter {\sc nmax}, a maximum number of bodies set by the user in the configuration file {\sc mercury.inc}. During execution of {\sc mercury}, the number of {\sc stat} elements used is equivalent to the actual number of bodies ({\sc nbod}) in the simulation. If {\sc stat} had been initialised then {\sc nmax} would need merely to be equal to or greater than {\sc nbod}. However, when using a compiler such as \textit{g77} or \textit{gfortran}, problems will be encountered unless {\sc nmax} is greater than {\sc nbod} by several hundreds.

Using a \textit{g77} compiler, simulations with 25 different initial conditions (Table~\ref{sims}), including the example given by {\sc Mercury}'s author \citep{Chamberssite}, were ran one or more times with different values of {\sc nmax}. An audit was then conducted of the number of bodies and a pattern emerged: simulations with a value of {\sc nmax} considerably higher than the number of bodies did not have problems; bodies disappeared from simulations with a comparatively small {\sc nmax} value.

\begin{figure*}
\centering
\epsfig{file=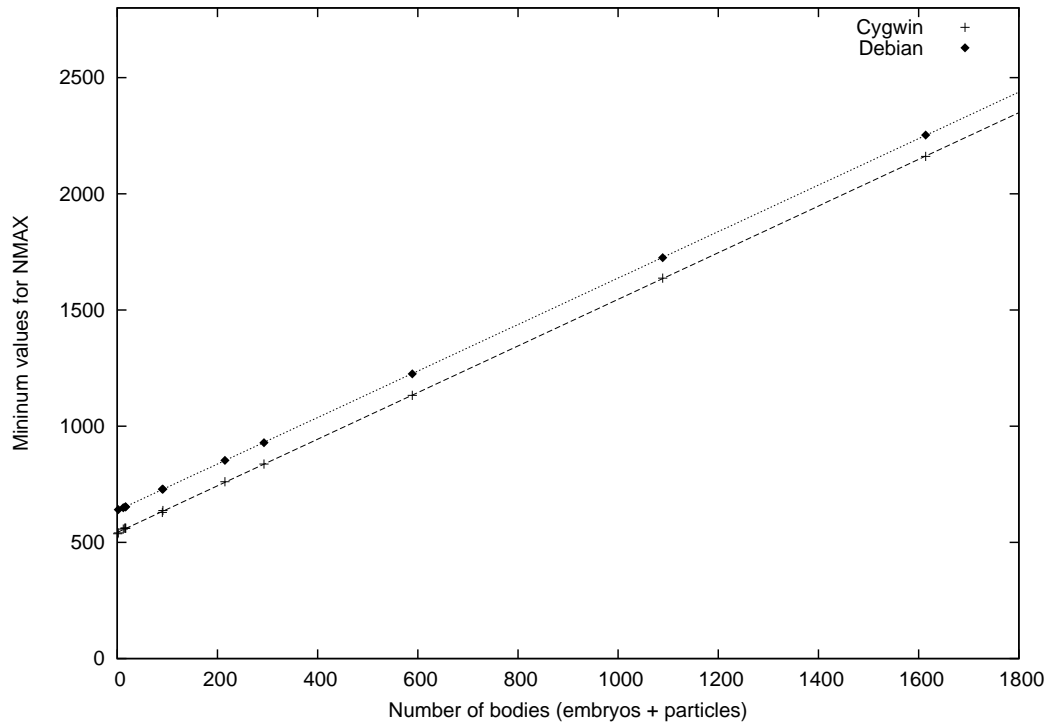,width=9.8cm,angle=270}
\caption{Minimum value of {\sc nmax} required for a problem-free execution as a function of number of bodies. The value depends somewhat on the environment; two were tested: Cygwin with GNU Bash, version 3.2.39(19)-release, and Debian 4.0 with kernel 2.6.18-5-amd64.}
\label{values}
\end{figure*}

A set of 12 simulations, with different numbers of bodies, were used to find the minimum limit of {\sc nmax} required for a problem-free execution (depicted with a $*$ in Table~\ref{sims}. Two different environments were used, a Cygwin with GNU Bash, version 3.2.39(19)-release, and a Debian 4.0 with kernel 2.6.18-5-amd64. For each execution, a short integration time (about 100 years) was used. The value of {\sc nmax} was varied in each simulation until the lower limit required to avoid losing bodies in a non-physical manner was found. The lower limit was found to be proportional to the number of bodies (Figure~\ref{values}). As the number of bodies tends to zero, the lower limit of {\sc nmax} tends toward a non-zero value around 500--700, the exact value depending on the computing environment.

\begin{table*}
 \centering
 \begin{minipage}{160mm}
  \centering
  \caption{Tests made to find discontinuities problem in Mercury's results. $^a$}
  \begin{tabular}{@{}cccccccc@{}}  
  \hline
	Big Bodies	&	Small Bodies	&	Giant Planets	&	Collisions	&	Central	&	Style	&	Algorithm	&	Times\\
  \hline
	0	&	1	&	0	&	no	&	Sun	&	Cartesian	&	Hybrid	&	x 1\\
$*$1	&	0	&	0	&	no	&	Sun	&	Cartesian	&	Hybrid	&	x 1\\
$*$1	&	1	&	0	&	no	&	Sun	&	Cartesian	&	MVS	&	x 1\\
	9	&	204	&	4	&	no	&	Sun	&	Cartesian	&	BS	&	x 3\\
	9	&	204	&	4	&	yes	&	Sun	&	Cartesian	&	Hybrid	&	x 1\\
	9	&	0	&	4	&	no	&	Sun	&	Cartesian	&	Hybrid	&	x 1\\
$*$11	&	1	&	0	&	no	&	Jupiter	&	Asteroidal	&	BS	&	x 2\\
	11	&	1	&	0	&	yes	&	Jupiter	&	Asteroidal	&	Hybrid	&	x 2\\
$*$11	&	4	&	0	&	yes	&	Jupiter	&	Asteroidal	&	Hybrid	&	x 1\\
$*$11	&	6	&	0	&	no	&	Jupiter	&	Asteroidal	&	Hybrid	&	x 1\\
	14	&	6	&	4	&	yes	&	Sun	&	Asteroidal	&	BS2	&	x 2\\
$*$14	&	2	&	4	&	yes	&	Sun	&	Asteroidal	&	BS2	&	x 1\\
	18	&	200	&	4	&	yes	&	Sun	&	Asteroidal	&	MVS	&	x 1\\
	68	&	204	&	1	&	yes	&	Sun	&	Asteroidal	&	Hybrid	&	x 2\\
	69	&	204	&	2	&	yes	&	Sun	&	Asteroidal	&	Hybrid	&	x 3\\
	72	&	204	&	1	&	yes	&	Sun	&	Asteroidal	&	Hybrid	&	x 1\\
	73	&	204	&	2	&	yes	&	Sun	&	Asteroidal	&	Hybrid	&	x 3\\
	88	&	204	&	1	&	yes	&	Sun	&	Asteroidal	&	Hybrid	&	x 3\\
$*$88	&	1525	&	1	&	yes	&	Sun	&	Asteroidal	&	Hybrid	&	x 2\\
	89	&	204	&	2	&	yes	&	Sun	&	Asteroidal	&	Hybrid	&	x 9\\
$*$89	&	500	&	2	&	yes	&	Sun	&	Asteroidal	&	Hybrid	&	x 1\\
$*$89	&	800	&	2	&	yes	&	Sun	&	Asteroidal	&	Hybrid	&	x 1\\
$*$89	&	1000	&	2	&	yes	&	Sun	&	Asteroidal	&	Hybrid	&	x 1\\
$*$89	&	1300	&	2	&	yes	&	Sun	&	Asteroidal	&	Hybrid	&	x 1\\
$*$89	&	1450	&	2	&	yes	&	Sun	&	Asteroidal	&	Hybrid	&	x 1\\
\hline
\label{sims}
\end{tabular}
\footnotetext{$^a$ Columns are: Number of big bodies, number of small bodies, number of giant planets, if accepts collisions or not, central body, style of data input, integrator algorithm and times the same instance of the simulation were simulated with small difference in bodies' positions. Simulations for Figure \ref{values} are shown by *.}
\end{minipage}
\end{table*}

The best fit for the points in the Cygwin environment is:

\begin{equation}
f(x) = 1.00341x + 542.708
\end{equation}
and in the Debian environment:

\begin{equation}
g(x) = 1.00009x + 637.329
\end{equation}
where x is the total number of bodies. These functions are basically the number of bodies plus an offset value. These tests offer a basic way to verify if old or recent simulations could be affected for the problem in {\sc stat} variable's values. We advise that any simulation with the value of the parameter {\sc nmax} close to these limits to have its results checked, if the compiler used was \textit{g77}, \textit{gfortran} or one with similar features.

\begin{figure*}
\centering
\epsfig{file=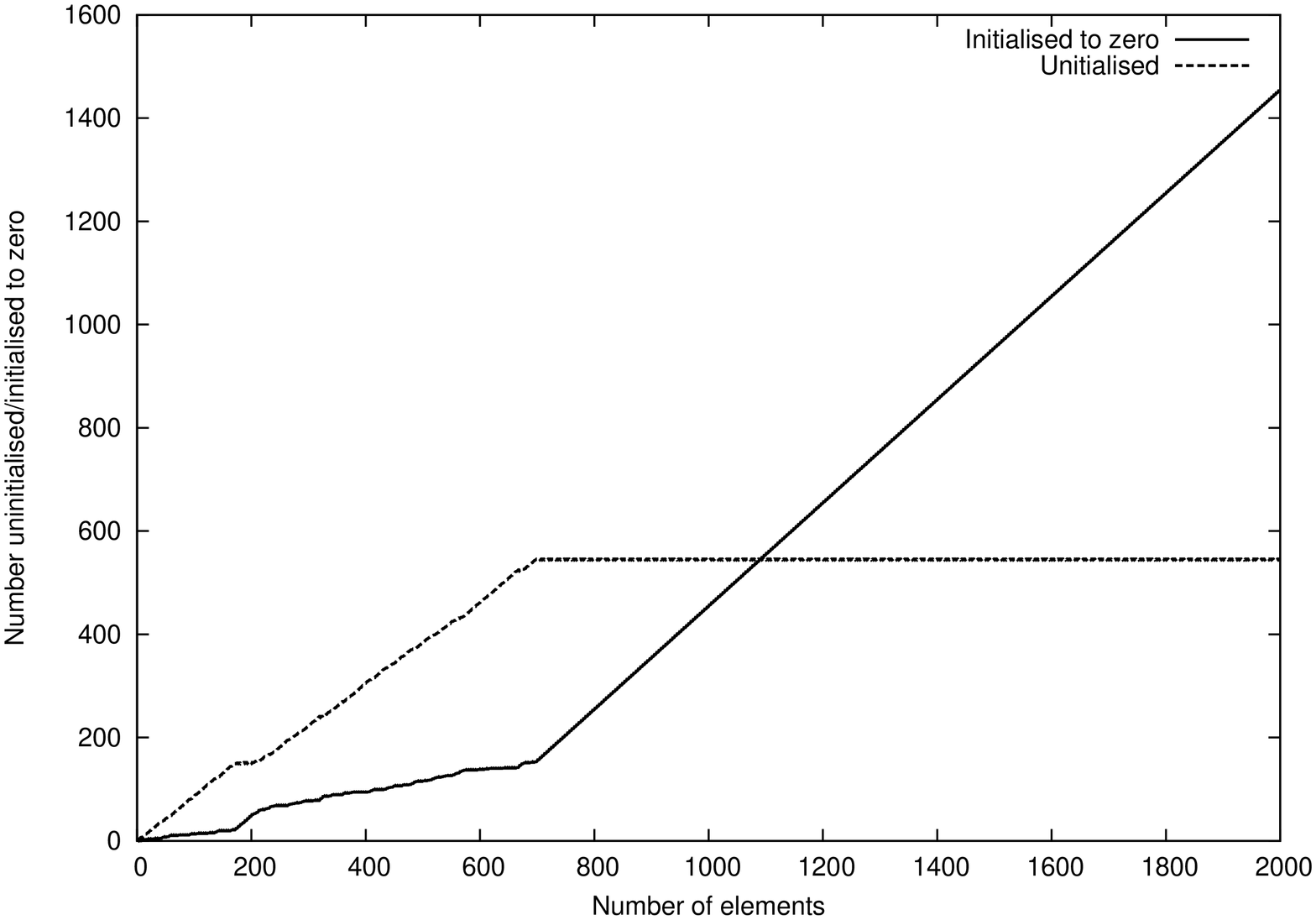,width=9.8cm,angle=270}
\caption{The number of elements that are and are not initialised to zero as a function of the number of elements, when using the \textit{g77} compiler under a Debian environment. As the array size increases (abscissa), the number of uninitialised elements increases steadily, with a smaller number of initialised elements spread throughout the array. When the number of elements passes 696, the number uninitialised remains constant and the number initialised increases linearly. Beyond this value, initialised elements are added at the beginning of the array, pushing the uninitialised elements to the end. The result is similar when using \textit{gfortran} but the turnoff occurs at 628 elements. If {\sc Mercury} has been compiled with such a compiler then the {\sc nmax} value would have to have been greater than the turnoff value plus the number of bodies in order for bodies not to have disappeared in a non-physical manner.}
\label{DPlot}
\end{figure*}

A simple test program was run, in which a variable array was declared but not initialised and its values output. The array values were mostly large positive, large negative and zero. Varying the size of the array, it was found that beyond a limiting size the values at the {\em beginning} of the array were 0. Figure~\ref{DPlot} shows this relation when using the \textit{g77} compiler in a 
Debian environment; when the number of elements exceeded 696, zeros began appearing at the beginning of the array. Therefore, if {\sc nmax} is greater than the sum of this minimum limit plus the number of bodies then no bodies will disappear non-physically.

The test was repeated with the \textit{gfortran} and \textit{ifort} compilers. With \textit{gfortran} the minimum limit was found to be 628. With \textit{ifort} the minimum value was found to be 0 as the compiler initialises all array elements to 0.

\section{Solution}
\label{solution}
Excepting when the used {\sc fortran} compiler is one that makes unconditional implicitly initialisation, a variable not explicitly initialised can have unpredictable behavior. In order to solve this problem, the variable array STAT must be initialised with zero in some point inside the file \textit{Mercury6\_2.for}, before the commands' block for the main calculations. We propose this initialisation to be done in the subroutine MIO\_IN, before (line 6046) or after (line 6111) the block of commands "Check for attempts to do incompatible things". Adding the command outside of the if statement (lines 5911-6045) for new and old integrations will guarantee it is going to be executed for both of them. The initialisation could be done with the three lines: 

\begin{tabbing}
do j=2, nbod\\
\hspace{3mm}STAT(j) = 0\\
end do\\
\end{tabbing}

With this command being executed each time an integration starts afresh or from dump files, the variable {\sc stat} will not receive random values, independently of the value of the parameter {\sc nmax} (it must still respect the basic rule: {\sc nmax} $\geq$ {\sc nbod}), the machine environment or the {\sc fortran} compiler. 

The initialisation could be done also in other points of the code, but one must be sure it is being done before the main calculations for both old and new integrations. 

\section{Conclusions}
\label{conclusions}

We repeated all the tests showed in Section \ref{charac} with a corrected version of {\sc Mercury}, initialising the variable array {\sc stat} at the end of the subroutine MIO\_IN in the source code. No discontinuities were seen in any results, independently of conditions. The new version is now being used in 40 new simulations of accretion dynamics. We believe this small change can improve the program making it more reliable for any type of N-body problem simulations. The corrected version can be downloaded from \url{http://www.astro.keele.ac.uk/~dra/mercury/}.

\section*{Acknowledgements}
{\small We are grateful to Sean Raymond for identifying an issue with simulations' results that led to the discovery of the bug.} 

\bsp~The MNRAS class file (\copyright Blacwell Science 2001) and the MNRAS bibliography style file were used in the preparation of this paper; they are available at \url{http://www.blackwellpublishing.com/static/mnras_latex.asp}.

\bibliographystyle{mn2e}
\bibliography{karlatorres}

\label{lastpage}
\end{document}